\documentclass[runningheads,citeauthoryear]{apinv}
\usepackage{epsfig,cite,graphics}
\usepackage{marvosym} %For astronomical symbols % % %
\usepackage[utf8]{inputenc}

\begin{document}

\title{Colours of the flickering source of Mira}
\titlerunning{Flickering of Mira}
\author{R. Zamanov\inst{1}, S. Boeva\inst{1}, B. Spassov\inst{1}, 
G. Latev\inst{1}, U. Wolter\inst{2}, K. A. Stoyanov\inst{1}}
\authorrunning{Zamanov et al.}
\tocauthor{R. Zamanov, S. Boeva, B. Spassov, G. Latev, U. Wolter, K. A. Stoyanov} 
% Command tocautor{} is used by the Latex to give author names 
% to the Contents of the volume (automatically generated)
\institute{Institute of Astronomy and NAO, Bulgarian Academy of Sciences, Tsarigradsko shose 72, 
           BG-1784 Sofia, Bulgaria  
	 \and  Hamburger Sternwarte, Universit\"at Hamburg, Gojenbergsweg 112, 21029 Hamburg, Germany 
	 \newline
	\email{rkz@astro.bas.bg;  kstoyanov@astro.bas.bg}    }
\papertype{Submitted on  8 March 2019; Accepted on  26 March 2019}	
% Papertype can be "Research report", "Review", "Invited lecture", "Conference talk", 
% "Conference poster", "Lecture at scientific seminar", "Summary of dissertation",  etc.
\maketitle

\begin{abstract}
We report  photometric observations in  Johnson  UBV bands of the short term 
variability of Mira. The amplitude detected is 0.16 mag in B band.  
Adopting interstellar extinction $E(B-V)\approx 0$, we find for the flickering source 
colour $B-V \approx 1.3$, temperature  $T \approx 3400$~K, and radius $R \approx 0.77$~R$_\odot$.
The colour of the flickering source is considerably redder than the average B-V colour of cataclysmic variables.
%We estimate colour of the flickering source  and temperature  of the flickering source. 
\end{abstract}

\section*{Introduction}
Mira (omicron Ceti) is  the prototype of the Mira-type variable stars.
This type stars are red giants in the very late stages of the stellar evolution, 
on the asymptotic giant branch (AGB), having pulsation periods of a few hundred days.
Mira pulsates with a period of 332 days (Hoffleit 1997) and an 
amplitude of about 7 magnitudes in V band
(see Fig.~\ref{f.AAVSO}). 
It provides an example of a binary system consisting of a mass-losing AGB star (Mira A) 
and a nearby (0.6 arcsec separation) companion Mira B, which is probably 
a white dwarf (Karovska et al. 1997, 2005; Sokoloski \& Bildsten 2010). 
% that will expel their outer envelopes 
% as planetary nebulae and become white dwarfs within a few million years. 
% Mira consists of a late type AGB giant and a hot component, which higly likely is a white dwarf.
% SIMBAD gives spectral type   M5-9IIIe+DA. 
Mira is classified as  S-type  symbiotic star in the recent catalogue based on the 2MASS, WISE and Gaia surveys 
(Akras et al. 2019), in other words it is an interacting, wide binary
system consisting of a red giant star that transfers matter to a much hotter companion. 
 
% usually a white dwarf or a neutron star. 
%  In the case of a white dwarfs as a primary, the sec-
%  This catalogue gives old and new infrared type = S, $T_{BB} = 2332 \; K$, 
%  $T_{cool}=3342 \; K$, $\lambda _{peak} = 1.24 \mu m $, no dust (Kiro ?).
%  GAIA DR2 distance = 
%  Light curve AAVSO 

Here we report CCD photometry of Mira and 
estimate colour, temperature and radius  of the flickering source. 

\section{Observations}

% ----------------------------------------------------------------------------- 

 \begin{figure*}   
%\mbox{}   
  \vspace{7.0cm} 
 \includegraphics{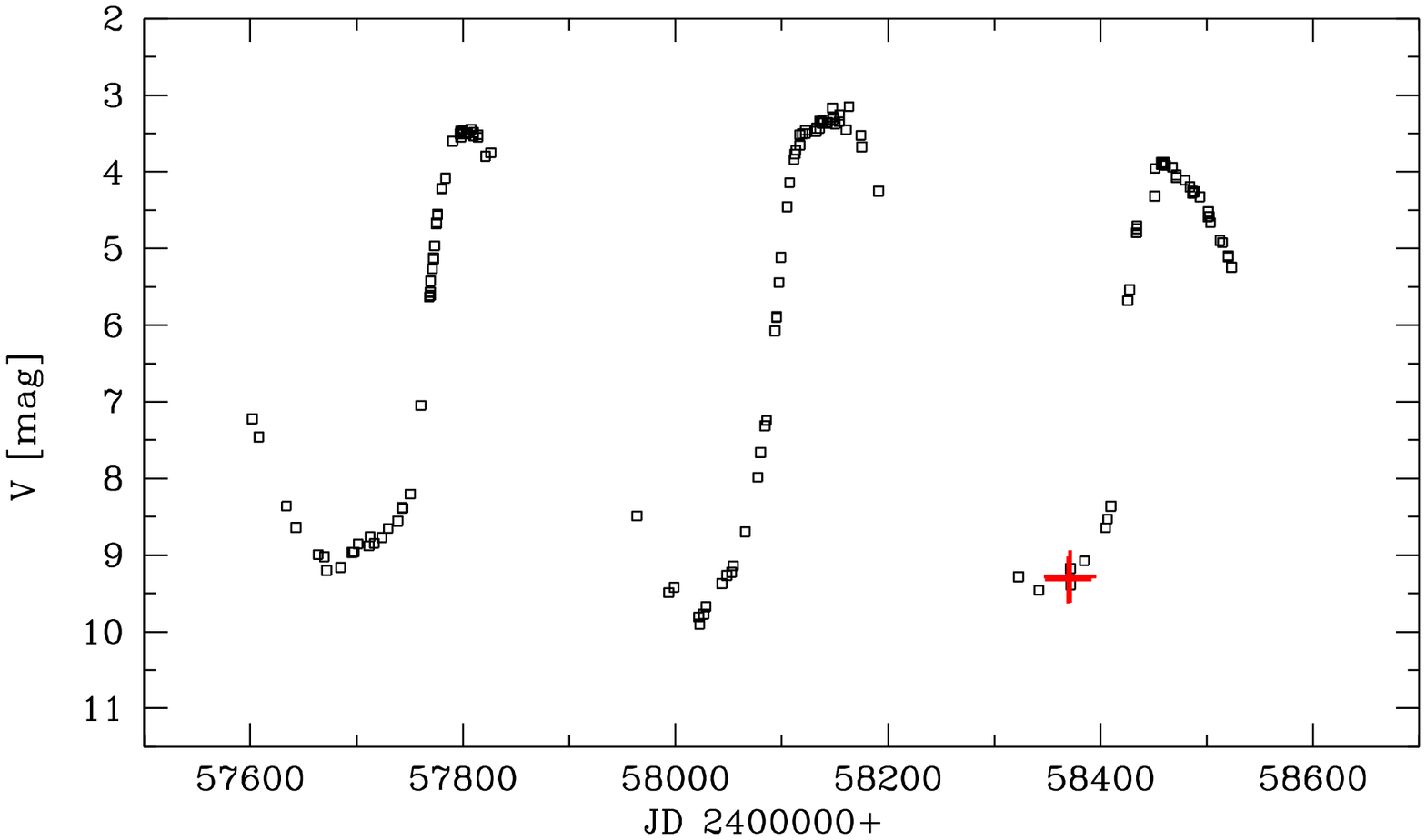} 
 \label{f.AAVSO}       
 \caption[]{Light curve of Mira in V band (AAVSO and British Astronomical Association Variable Star Section
 data). The two red crosses indicate our V band runs. } 
  \vspace{11.5cm} 
  \includegraphics{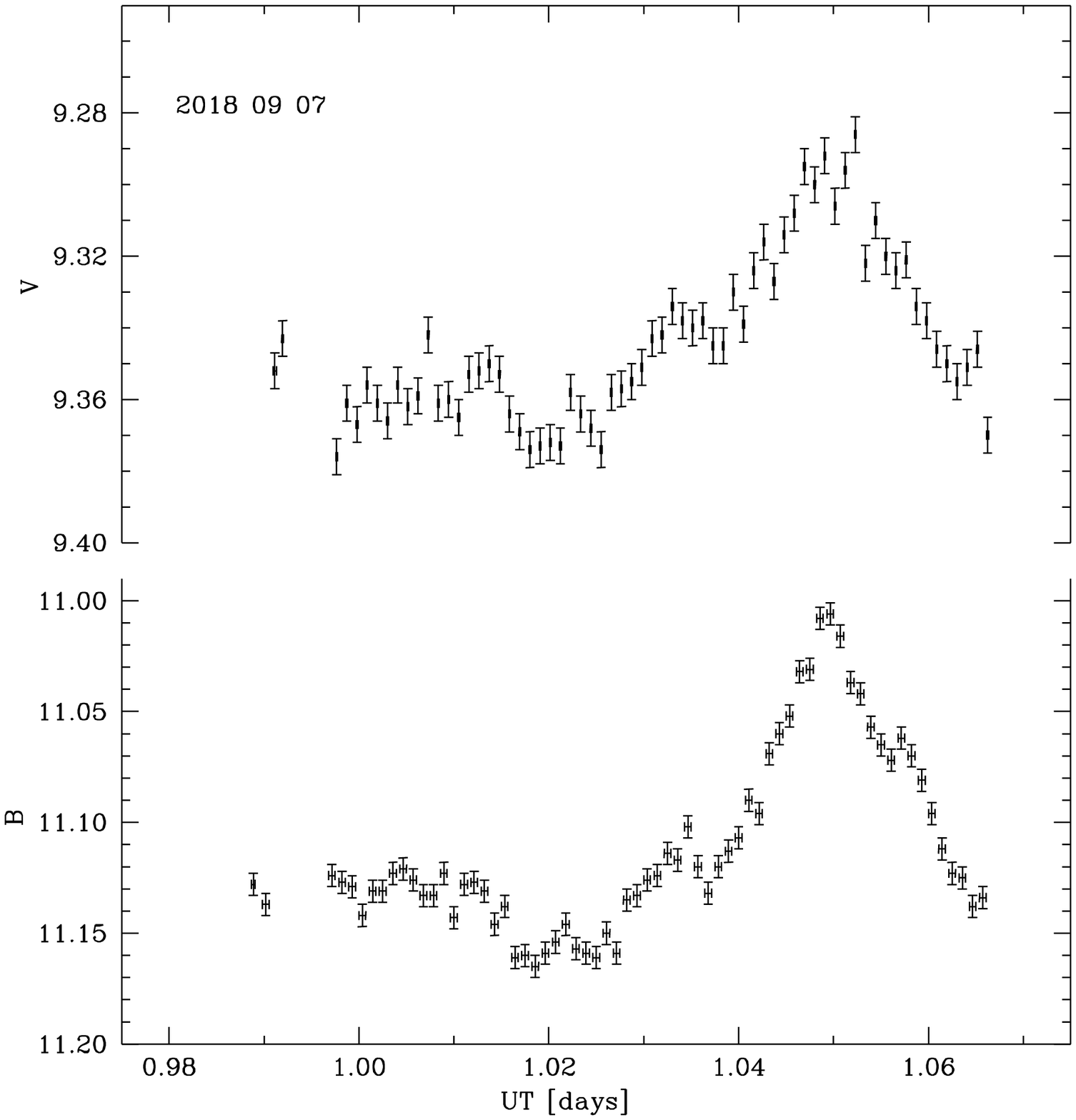}      
  \includegraphics{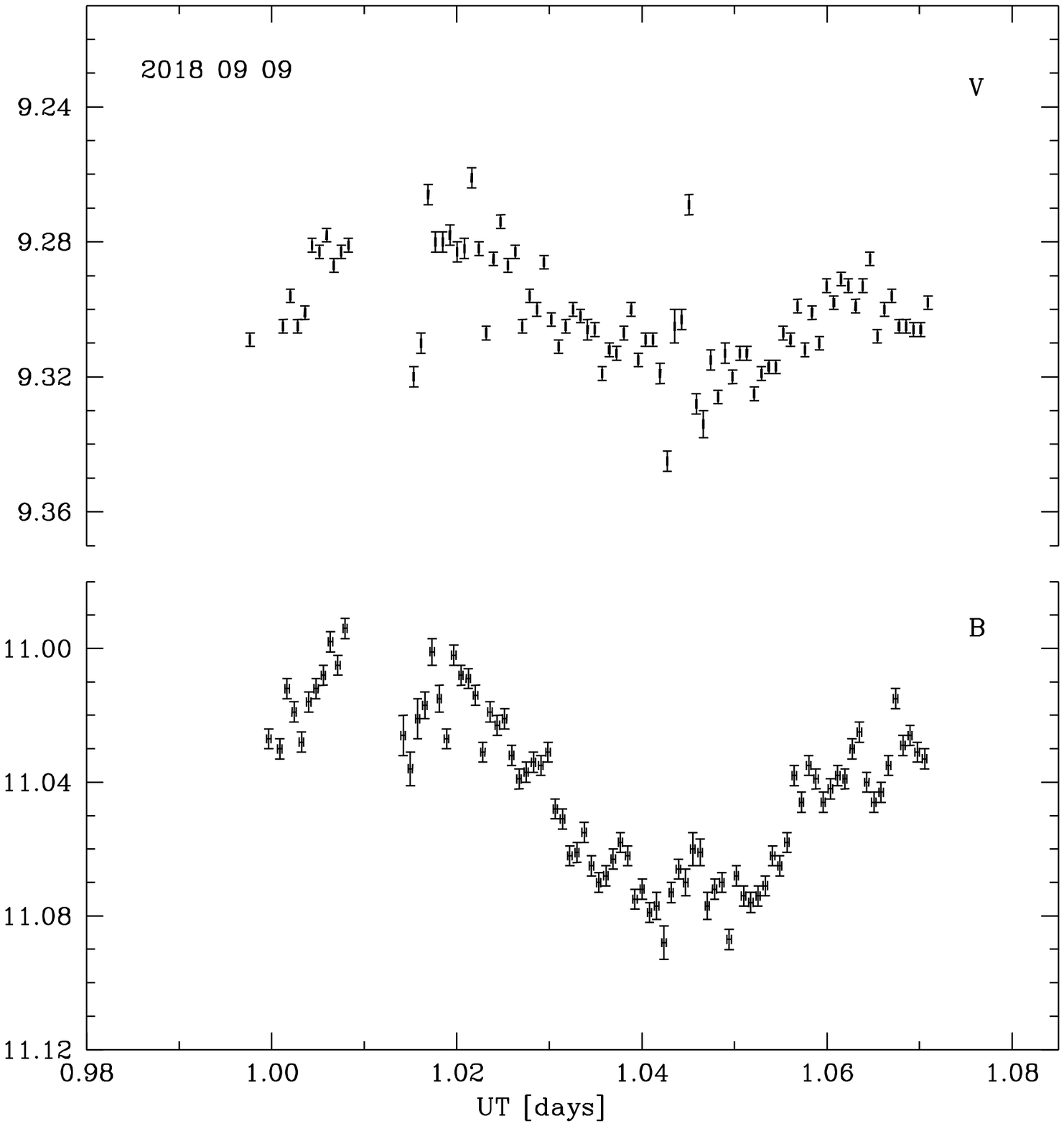}      
  \includegraphics{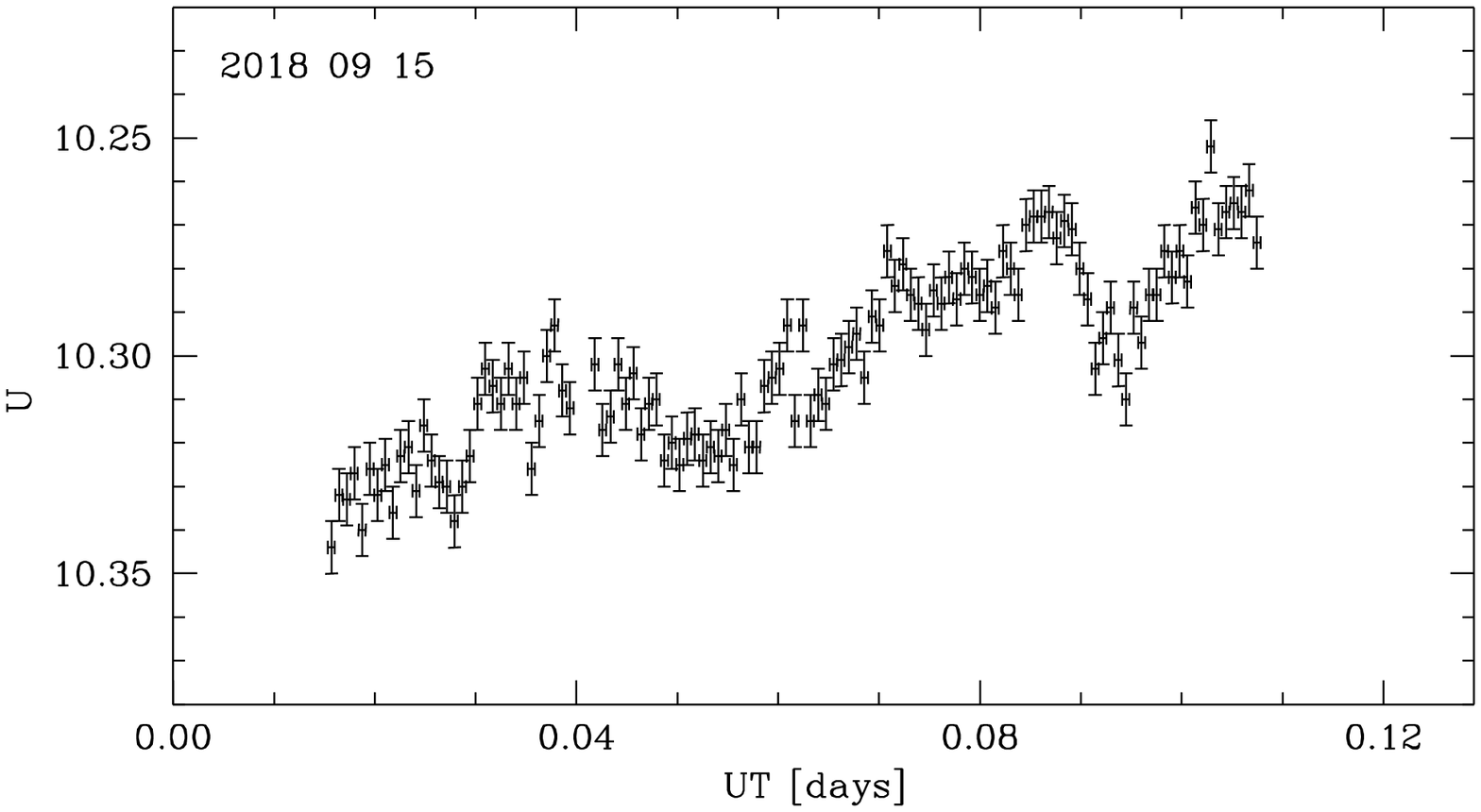}      
\caption[]{Short term variability of Mira.
  The date of observations is in format YYYYMMDD.
}  
\label{f.our}      
\end{figure*}        
%------------------------------------------------------------------------------

%---------------------------------------------------------- 
\setlength{\tabcolsep}{5pt}
\begin{table}
  \begin{center}
  \caption{Journal of the CCD observations of Mira. 
  The columns give  date of observation (in format YYYYMMDD), the telescope used, band, 
  number of  data points in the run and exposure time,  UT-start and UT-end of the run. }
  \begin{tabular}{l l | c  c | c l r r }
 date     & telescope & band  &  $N_{pts}$       & UT start - end      &  \\
          &           &       &                  & hh:mm -- hh:mm      &  \\
 \hline
          &                   &   &                                       \\
 20180907 & 60cm Belogradchik & B & 67 x 60 sec  & 23:44 - 01:34 &     &  \\ 
          &                   & V & 67 x 15 sec  & 23:47 - 01:35 &     &  \\
          &                   &   &                                       \\	  
 20180909 & 60cm Belogradchik & B & 81 x 40 sec  & 23:57 - 01:42 &     &  \\  
          &                   & V & 83 x 10 sec  & 23:57 - 01:42 &     &  \\
          &                   &   &                                       \\
 20180915 & 50/70cm Schmidt   & U & 119 x 60 sec & 00:22 - 02:34 &     &  \\  
          &                   &   &                                       \\
 \label{tab1}
 \end{tabular}
 \end{center}
 %-----------------------------------------------------
  \begin{center}
  \caption{Photometry of Mira.  In the table are given: date of observation, band, 
minimum, maximum and average magnitudes  in the corresponding band, standard
deviation of the mean, typical observational error.
}
  \begin{tabular}{l c | r  r r  r r rrr}
 date     &  band     &  min  &  max  & average & stdev  &  merr  &  \\
          &           & [mag] & [mag] &  [mag]  &  [mag] &  [mag] &  \\
 \hline
          &   &         &         &          &        &        & \\
20180907  & B &  11.006 &  11.165 & 11.1115  &  0.041 & 0.004  & \\  
          & V &   9.286 &   9.376 &  9.344   &  0.023 & 0.003  & \\   
          &   &         &         &          &        &        & \\	  
20180909  & B &  10.998 & 11.088  & 11.0427  &  0.023 & 0.004  & \\
          & V &   9.261 &  9.345  &  9.3022  &  0.016 & 0.003  & \\
          &   &         &         &          &        &        & \\
20180915  & U & 10.2523 & 10.3440 & 10.3008  & 0.0210 & 0.006  & \\
          &   &         &         &          &        &        & \\
 \label{tab2}
 \end{tabular}
 \end{center}
 %--------------------------------------------
  \begin{center}
  \caption{Estimated parameters of the average flickering source of Mira for
  E(B-V)=0,  E(B-V)=0.04 and  E(B-V)=0.70. 
   }
  \begin{tabular}{l c | r  r r  r r rrr}
  
 date     &  parameter      & E(B-V)=0  & E(B-V)=0.04  & E(B-V)=0.70  &  \\
          &                 &           &              &              &  \\
 \hline
          &                 &           &              &       &        \\
 20180907 & B-V             & 1.22      & 1.18         & 0.52  &        \\ 
          & T [K]           & 3556      & 3896         & 6375  &        \\
          & R [R$_\odot$]   & 0.66      & 0.65         & 0.45  &        \\	  
          &                 &           &                               \\
 20180909 &  B-V	    & 1.36      & 1.32         & 0.66  &        \\
          &  T [K]     	    & 3253      & 3604         & 5636  &        \\
          &  R [R$_\odot$]  & 0.89	& 0.87         & 0.60  &        \\
&  &   & \\
\hline 
\\
 \label{tab3}
 \end{tabular}
 \end{center} 
\end{table}
%-------------------------------------------------

The observations were performed with two telescopes equipped with CCD cameras:
the 60 cm telescope of the Belogradchik Observatory, Bulgaria (Strigachev \& Bachev 2011)  
and 
the 50/70 cm Schmidt telescope of the Rozhen National Astronomical Observatory, Bulgaria.   
Two comparison stars were used during the data reduction. Both are taken from AAVSO data base:   
HD 14411 (U=12.501 B=10.794 V=9.330)
and 
the star located 
at  J2000 coordinates  RA=02:19:08.51 Dec=-02:54:51.0 (U=14.754 B=14.471 V=13.740).

% st1 for which we adopt   M4 V  B=10.90  V=9.31  B-V=1.59  U-B=1.15  U=12.05. 

The journal of observations is given in Table~\ref{tab1} and the light curves are plotted on Fig.~\ref{f.our}. 
The detected amplitude in  V band is $\approx 0.1 mag$,  in  B band is 0.16 mag,  and  in U band is 0.09 mag. 
In Table~\ref{tab2} are given the measured magnitudes in each run.
Fig.~\ref{f.AAVSO} illustrates that
    (1) Mira's V-band brightness varies during the last 1000 days;
    (2) our V band observations are in agreement with other data
and (3) our runs  are close to the minimum of the Mira cycle.

\section{Flickering source}

% Bruch (1992) proposed that the light curve of
% a white dwarf with flickering  can be separated into two parts -- constant light
% and variable (flickering) source. 

Following  Bruch (1992), we separate the light curve  into two parts -- constant light
and variable (flickering) source. 
We calculate the flux of the flickering light source 
as $F_{\rm fl1}=F_{\rm av}-F_{\rm min}$, where $F_{\rm av}$ is the average flux 
during the run and $F_{\rm min}$ is the minimum flux during the run
(corrected for the typical error of the observations).
An extension of the method is proposed by Nelson et al. (2011), 
who suggests to use the $F_{\rm fl2}=F_{\rm max}-F_{\rm min}$, where $F_{\rm max}$ 
is the maximum flux during the run. 
Practically, the method of Bruch (1992) refers to the average luminosity of the flickering source, 
while that of Nelson et al. (2011) -- to its maximal luminosity. 
$F_{\rm fl1}$ and $F_{\rm fl2}$  have been calculated for each band, using the values 
given in Table~1 and 
the calibration for a zero magnitude star $ F_0 (B) =6.293 \times 10^{-9}$  erg cm$^{-2}$ s$^{-1}$ \AA$^{-1}$,   
$\lambda_{eff}(B)=4378.12$~\AA, 
$F_0 (V) = 3.575 \times 10^{-9}$  erg cm$^{-2}$ s$^{-1}$ \AA$^{-1}$ and  $\lambda_{eff}(V)=5466.11$~\AA\ 
as given in the Spanish virtual observatory 
Filter Profile Service  (Rodrigo et al. 2018, see also Bessel 1979).  

$Hipparcos$ gives parallax of Mira  $10.91 \pm 1.22$ milliarcseconds, which corresponds to 
distance $d=92$~pc   (No GAIA paralax is available for Mira). 
Mira is very close to the Earth and the interstellar extinction is probably $E(B-V) \approx 0$.
However the neutral hydrogen absorbing column
 $N_H = 2.24 \times 10 ^{20}$~cm$^{-2}$  is derived from UV spectral modeling
(Wood et al. 2002). From model fitting of  ROSAT and XMM-Newton
X-ray data, Kastner \& Soker (2004)  find  $ 2.0 \times 10^{21}$ cm$^{-2}$ $< N_H  < 4.5 \times  10^{21}$~cm$^{-2}$. 
This X-ray derived $N_H$ is a factor of $>20$ larger than the neutral H absorbing column determined
from analysis of the HI Ly$\alpha$ line (Wood et al. 2002). 
This discrepancy suggests that the UV and X-ray emission arise in
different zones around Mira B and, specifically, that the X-ray 
emitting region may be embedded within accretion streams
that effectively attenuate the X-rays, as described by Jura \& Helfand (1984).
Following the relation between $N_H$ and reddening  $N_H/A_V = 1.79 \times 10^{21}$ cm$^{-2}$ mag$^{-1}$
(Predehl \& Schmitt 1995), 
we estimate that  $N_H = 2.24 \times 10 ^{20}$~cm$^{-2}$ corresponds to E(B-V)=0.04, 
$N_H = 2 \times 10^{21}$~cm$^{-2}$ -- to  E(B-V)=0.35, 
and $N_H = 4.5 \times 10^{21}$~cm$^{-2}$ -- to    $E(B-V)= 0.79$.

In Table~\ref{tab3}  are given the parameters of the flickering source,  calculated
using $F_{av}$ and following Bruch (1992). 
Assuming $E(B-V) \approx 0$, we estimate colour 
and temperature  of the flickering source  $B-V = 1.29 \pm 0.10 $  and  $T = 3400 \pm 200$~K
-- see Table~\ref{tab3} for more details. 
Using the method of Nelson et al. (2011) and  $E(B-V) = 0$,  we estimate similar colour and temperature:  

 20180907 \hskip 0.5cm   $B-V =1.14 $,  $T=3979$~K,  $R=0.97$~R$_\odot$;  

 20180909 \hskip 0.5cm   $B-V =1.52 $,  $T=3250$~K,  $R=1.88$~R$_\odot$. \\
As can be expected the size is larger, because this method  refers to the maximum of the flickering.

%------------------------------------------------- 
%\setlength{\tabcolsep}{5pt}
%\begin{table}
%\end{table}
%-------------------------------------------------

\section{Discussion and Conclusion}

Mira~AB is a very wide binary system. The infrared spectroscopy  (Hinkle et al. 2013)  detected no orbital motion 
in agreement with previous estimates
of the orbital period $\ge 100$ yr and semimajor axes $\sim 50$~au. 
% ESPERO - Mira.20181029.fits 
%Roche lobe overflow is thermally unstable when the donor star in a binary is an evolved
%giant with a mass in excess of 5 $M_{wd}$ /6 (where M$_{wd}$ is the mass of the accreting white dwarf 
%(Webbink, Rappaport, \& Savonije 1983). 
ALMA observations revealed  a mass of gas surrounding Mira B, 
with a size of a few tens of AU, which is interpreted as gas flowing from Mira A toward Mira B
(Nhung et al. 2016). Hence the white dwarf in Mira  accretes  material via gravitational capture of the red-giant
wind. 

Sokoloski \& Bildsten (2010) detected optical flickering from Mira with amplitude 0.15 - 0.30 mag 
in B band during 5 runs in 1997 - 1998. 
Snaid et al. (2018) found flickering with amplitude  0.05 mag in g' band  in September 2015.
The amplitude of the flickering in our data is similar to these results.
Given the discussion in Sect.~6.2 of Sokoloski, Bildsten \& Ho (2001) about the difficulty of
producing rapid variability from the nebular emission, we 
expect  the rapidly variable component of Mira 
to reflect the physical origin of the variations in the accretion disc around white dwarf.

For the flickering source of  the symbiotic Mira EF Aql, 
Zamanov et al. 2017 calculated  $(B-V)= 0.35 \pm 0.05$, which is considerably more blue than the colour of the 
flickering source of omi Cet.  
The B-V colour of the flickering source is calculated in a number of cataclysmic variables and two recurrent novae 
(Bruch 1992; Zamanov et al. 2015). It is in the range $-0.19 \le B-V \le 1.09$, 
with average value of 0.16, median value of 0.11, and standard deviation of 0.24. 
Our results for Mira indicate that if the interstellar extinction is 
low, its flickering source is considerably 
more red than the average colour of the cataclysmic variables. 
Possible reasons can be: \\
 -- (1) the flickering source of Mira has lower temperature than the flickering source 
in cataclysmic variables; \\
 -- (2) the flickering of  Mira  is coming from 
 almost the same place, where the X-rays are generated and 
 the flickering   source is 
embedded within the accretion flow 
(like the X-ray source, see Kastner \& Soker 2004) which makes it to appear redder.

 %
 % 1995A\&A...293..889P	 Predehl, P.; Schmitt, J. H. M. M.	
 % X-raying the interstellar medium: ROSAT observations of dust scattering halos.
 % $N_H / A_V = 1.79 \times 10^{21}$.
 %
 % \section{Conclusion}
 % Our result point that the flickering source of Mira is very. Possible reasons  

\vskip 0.5cm

{\bf Acknowledgements: }  This work is supported by the grant K$\Pi$-06-H28/2 08.12.2018
(Bulgarian National Science Fund).
We acknowledge with thanks the variable star observations from the AAVSO
International Database and the BAA Photometry Database 
contributed by observers worldwide and used in this
research.

%\newpage

\end{document}